\documentclass[twocolumn,showpacs,preprintnumbers,amsmath,amssymb]{revtex4}

\usepackage{graphicx}
\usepackage{epsfig}

\usepackage{txfonts}

\newcommand{\be}{\begin{eqnarray}}
\newcommand{\ee}{\end{eqnarray}}
\newcommand{\sbe}{\begin{eqnarray}}
\newcommand{\see}{\end{eqnarray}}
\newcommand{\pslash}{\not\!p}

\begin{document}
\preprint{\bf{JLAB-THY-10-1295}}
\title{{\Large\bf{Strong Coupling Continuum QCD}}}
\author{M.R. Pennington}
\affiliation{Theory Center, Thomas Jefferson National Accelerator Facility\\ 12000 Jefferson Avenue, Newport News, VA 23606, USA}
\date{November 2010}
\begin{abstract}
The Schwinger-Dyson, Bethe-Salpeter system of equations are the link between  coloured quarks and gluons, and colourless hadrons and their properties. This talk reviews some aspects of these studies from the infrared behaviour of ghosts to the prediction of electromagnetic form-factors. 

\end{abstract}


\maketitle


\section{Fermions, bosons in the continuum}
Strong coupling QCD in the continuum is the physics of the real world of  gluons, ghosts and quarks: a world in which quarks have their physical masses and the pion at 140 MeV is by far the lightest hadron.  This is the world that can be studied using the field equations of the theory and the consequent bound state equations: the Schwinger-Dyson  and Bethe-Salpeter equations~\cite{roberts,tandy,fischer-rev}. These constitute an infinite set of nested integral equations that we cannot solve except in rather particular truncations. The best known approximation is of course that of perturbation theory, where every Green's function is expanded in powers of the coupling and we stop at some finite order. However, the key physics that builds hadrons and determines their properties is one of strong coupling. Nevertheless, the beauty of  the perturbative expansion of a gauge theory, whether QED or QCD, is that the Green's functions  are multiplicatively renormalizable and fulfil the 
 requirements of gauge invariance at every order of truncation. Indeed, these are key properties of the full theory for any strength of coupling. They thus serve as guides for how we might usefully truncate a strong coupling expansion.

To start the strong coupling study of bound states one needs to know their basic building blocks, the propagators for quarks, gluons and ghosts and of course their interactions. The fact that the field equations are an infinite set of  coupled equations may make one think that even to solve for the propagators (the two-point functions) one needs to know not just the three and four point interactions, but even the 24-point function and beyond. If that were the case we could hardly make progress. While the fermion and boson propagators depend on two functions that multiply the two independent spinors and tensors, respectively, the fermion-boson interaction depends on 12 functions and the 10-point function some huge number. This looks impossible.  Fortunately, the two point functions do not depend on each of these independently, but only collectively.   The properties of multiplicative renormalizability and gauge invariance of the two point functions pull through to these Green'
 s functions the key elements of the higher point functions.
 
\begin{figure}[h]
\vspace{2mm}
  \includegraphics[width=0.9\columnwidth]{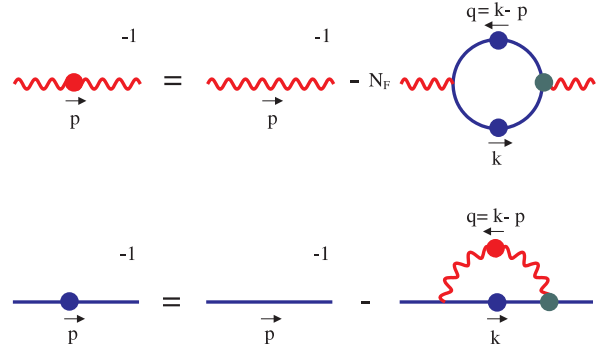}
  \caption{Schwinger-Dyson equations for the inverse boson and fermion propagators in QED. The dots denoted fully dressed quantities}
\vspace{2mm}
\end{figure} 
In general, the inverse boson propagator carrying momentum $p$, as shown in Fig.~1 for QED, has two independent tensor structures, so
\be
\Pi^{\mu\nu}(p)\;=\; A(p^2)\,p^2\,g^{\mu\nu}\;-\;B(p^2)\,p^\mu p^\nu
\ee
However in a covariant gauge the renormalization functions $A$ and $B$ are not independent. Indeed, in the Landau gauge $A(p^2)\,=\,B(p^2)$. For the boson propagator in QED, two conditions must be imposed to ensure this. The first is that the full fermion-boson vertex satisfies the Ward-Green-Takahashi (WGT) identity. To see how this works~\cite{mexico} consider the Schwinger-Dyson equation for the inverse photon propagator, Fig.~1:
\sbe
&&\hspace{-10mm}\Pi^{\mu\nu}(p)\;=\;\Pi^{\mu\nu}_0(p)\\\nonumber
&&+ \frac{g^2}{(2\pi)^4} \int d^4k \; {\rm{Tr}}\,\left[\gamma^\mu\,S_F(k)\,\Gamma^\nu(k,q)\,S_F(q)\right]
\see
where $\,q\,=\,k-p\,$.
We contract the propagator with $p_\mu p_\nu$, and implementing the WGT identity
\be
q_\mu\,\Gamma^\mu(k,p)\;=\;S_F^{\,-1}(k)\,-\,S_F^{\,-1}(p)\quad 
\ee
yields
\be
p_\mu \Pi^{\mu\nu}p_\nu = \frac{g^2}{(2\pi)^4} \int d^4k\; {\rm{Tr}} \left[\pslash \left(S_F(k-p) - S_F(k)\right)\right] 
\ee
since the bare propagator is transverse in the Landau gauge. The loop integral in Eq.~(4) runs over all momentum components from $-\infty$ to $+\infty$. If the integral were convergent, it would be obvious shifting the first term so $k-p \to k$ that on integration the answer would be zero. However, with a cut-off regulator this is no longer the case. We have to ensure that the integrals are  regulated in a translationally invariant way if the integral is to vanish. Thus one must not just satisfy the WGT identity, but to be able to  make the infinite integrals finite impose multiplicative renormalizablility too. 

It is in fact only very recently that we have learnt how to construct a full fermion-boson vertex that ensures both the fermion and boson propagators in QED are multiplicatively renormalizable~\cite{kizilersu}.  
An illustration of this is shown in Fig.~2 for the photon renormalization function~\cite{kiz_numerics}. (The boson renormalization functions plotted here, and later for QCD, are all defined as the propagator multiplied by $p^2$.) 
In Fig.~2 we see that while with a bare vertex, or even the Ball-Chiu version~\cite{ball-chiu}, this function in an Abelian theory is strongly gauge dependent, with the Kizilers\"u-P vertex~\cite{kizilersu} this is not the case. Since in QED the physical coupling is proportional to this renormalization function, its independence of the gauge is a clear necessity. 
\begin{figure}[h]
\includegraphics[width=0.98\columnwidth]{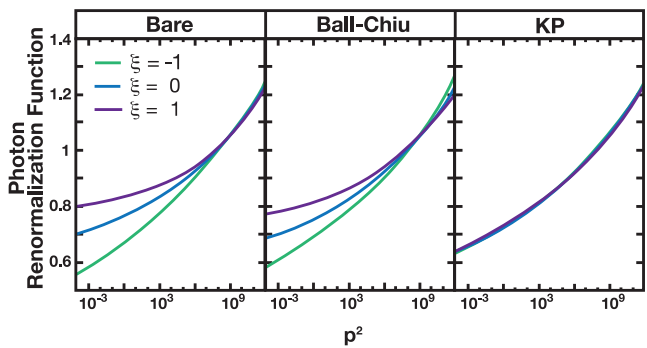}
  \caption{Log-log plot of the momentum dependence of the photon renormalization function in different covariant gauges defined by $\xi=-1, 0, 1$. These are the solutions of the coupled equations shown in Fig.~1 with $\alpha =g^2/4\pi =0.2$ at $p=10^4$, for three different ansatze for the fermion-boson vertex: (i) bare, (ii) Ball-Chiu~\cite{ball-chiu}, (iii) KP~\cite{kizilersu}. Only in the latter case is the result essentially gauge independent~\cite{kiz_numerics}.}
\end{figure}

\section{QCD in the continuum}
Studies of the gauge sector of QCD are more complicated. It was Baker, Ball and Zachariasen (BBZ)~\cite{bbz} who first used the axial gauge to investigate the behaviour of the gluon propagator. They understood the importance of constructing the triple gluon vertex to satisfy the relevant Slavnov-Taylor identity --- the non-Abelian extension of the WGT identity of Eq,~(3). Axial gauges seem at first sight ideal as they have no ghosts and only transverse gluons. However, one never gets anything for nothing. In such gauges, the gluon propagator depends on two independent functions: coefficients of two independent transverse tensors that can be constructed from $g^{\mu\nu}$, the gluon momentum $p^\mu$ and the axial vector $n^\mu$. At lowest order in perturbation theory, one of these functions is unity, and the other is zero. BBZ~\cite{bbz} assumed that this latter function remained zero even non-perturbatively, and deduced that the gluon propagator behaved like $1/p^4$ at infrared momenta. With the potential between infinitely heavy quarks controlled by one gluon exchange, this generates a linearly rising "confining" vector potential. Unfortunately, West~\cite{west} showed that in axial gauges the gluon could be no more singular than $1/p^2$ and both gluon functions control the infrared behaviour. Attempts to solve the coupled system for these have so far failed largely because of the difficulty of dealing consistently with the $\,n\cdot p\,$ singularities non-perturbatively.

 Consequently, attention turned to covariant gauges, in particular that of Landau, where of course one has to deal with ghosts. In early numerical studies, for instance that by Nick Brown and myself~\cite{bp}, the ghosts were treated perturbatively, just to ensure the ultraviolet behaviour of the gluons was correct.  However, it was von Smekal and Alkofer~\cite{smekal}, who showed that ghosts were essential in the infrared too. Indeed, they found that the ghosts were singular in the infrared, while the gluons were finite or vanishing. They showed these to be closely correlated in what is called the {\em scaling} solution, with the ghost behaving like $p^{-2\kappa-2}$, while the gluon propagator is $p^{\,4\kappa-2}$ as the momentum $p \to 0$, with $\kappa \simeq 0.6$. The renormalization (or dressing) functions (which recall are the propagators multiplied by $p^2$) are sketched in Fig.~3.

For much of hadronic physics the exact behaviour of the gluon and ghost propagators in the deep infrared is not relevant, so we postpone discussion of this till later and instead turn to their effect on quarks and then on the hadrons they build. In the world of light hadrons, the {\em up} and {\em down} quarks have current masses that are very much less than the scale of $\Lambda_{QCD}$. Even in Abelian theories dynamical mass generation can readily occur, provided the interactions are strong enough, typically $\alpha \simeq \pi/3 \simeq O(1)$. While perturbatively the vacuum is almost empty, strong long-range interactions change its nature. Then particles that appear in the Lagrangian with no mass propagate through this medium like heavyweights. Since chiral symmetry breaking is such a key feature of the low energy strong interaction, modelling this has been a much studied aspect of QCD~\cite{maris-tandy,roberts2}. The  Schwinger-Dyson equation (SDE) for the fermion propagator is shown in Fig.~1. 
As already remarked, this depends on two functions, the mass function ${M}(p^2)$ and the wavefunction renormalization ${F}(p^2)$, such that the full propagator is 
\be 
S_F(p)\;=\;\frac{{F}(p^2)}{\pslash\,-\,{M}(p^2))}\quad ,
\ee
\noindent where for the bare quantities ${F} =1$ and ${M} =m_0$. 
\begin{figure}[h]
\vspace{3mm}
  \includegraphics[width=0.9\columnwidth]{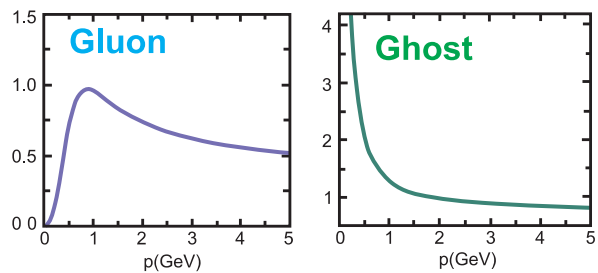}
  \caption{Gluon and ghost renormalization functions for the \emph{scaling} solution as found by \cite{smekal}.}
\end{figure}

In QCD, one can study these quark functions alone by modelling the product of the dressed gluon propagator and the quark-gluon vertex (lower graph in Fig.~1), as shown by Maris and Tandy~\cite{maris-tandy}. Regardless of whether the gluon is enhanced in the infrared or suppressed, the key region for physics is the momentum region of $p \simeq \Lambda_{QCD}$ (with calculations usually performed in a momentum subtraction scheme). There the effective strength is enhanced, the coupling becomes larger than 1 and even if $m_0=0$ a mass is generated. Indeed, one finds that for current masses for the {\em up} and {\em down} quarks defined in the perturbative regime to be just a few MeV, a mass of 300-400 MeV in the low momentum region is naturally generated, as seen in Fig.~4~\cite{williams,MR}. Since the SDEs in the continuum can be studied for any value of the current mass, one can readily increase these to 30-100 MeV and so compare with lattice results~\cite{bowman}. As shown in~\cite{roberts3}, these agree remarkably well. One of the benefits of working in the continuum is that one can also set the quark mass to zero. Then the behaviour of the quark propagators differs just a tiny bit from that with 3-5 MeV (only on a log-log plot like Fig.~4 is this difference to be seen). However, in the massless case one can apply the operator product expansion to the large momentum behaviour of the quark mass function of Fig.~4 and learn that is controlled by a ${\overline q}q$ condensate of $-(250$ MeV)$^3$, where the scale is essentially set by $\Lambda_{QCD}$. Thus it is  strong long range correlations between quarks and antiquarks that largely breaks chiral symmetry. Moreover,  this size of ${\overline q}q$ condensate agrees  with the value determined by precision studies of low energy $\pi\pi$ final state interactions from BNL-E865~\cite{e865} and the recent CERN-NA48~\cite{na48}, and through the Gell-Mann-Oakes-Renner relation~\cite{gmor} contributes $>90\%$ to the physical pion mass~\cite{colangelo}.

\begin{figure}
  \includegraphics[width=0.98\columnwidth]{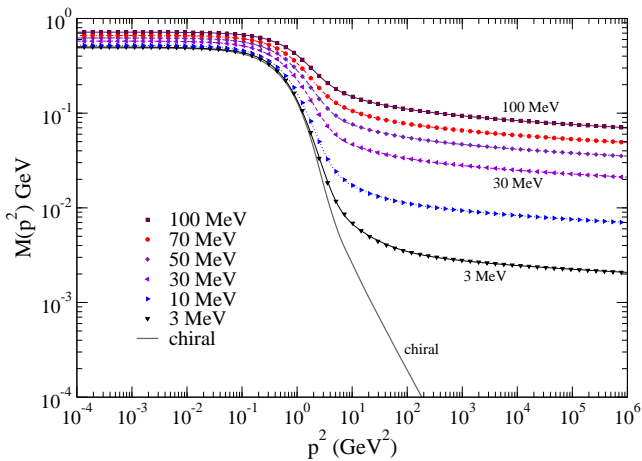}
\vspace{2mm}
  \caption{Log-log plot of the momentum dependence of the mass function $M(p^2)$ of Eq.~(5) for different values of the current mass of the quark that label the lines. The chiral limit of zero current mass is also shown.      These results
 from Ref.~\cite{williams} are essentially the same as those obtained earlier by Maris and Roberts~\cite{MR} covering a bigger range of masses required for 5 flavours, relevant to the discussion of Fig.~5.}
\vspace{3mm}
\end{figure}

\vspace{2mm}
This light quark propagator interpolating from small ultraviolet current masses to constituent masses below 500 MeV is an essential building block for hadronic bound states. Coupling the Schwinger-Dyson equations with those of Bethe-Salpeter allows both the static and dynamical properties of light flavoured hadrons to be calculated.
This provides a framework for studying the relation between hadronic quantities, for instance, the masses of the lightest pseudoscalar and the lightest vector meson. Such studies show the $\rho$-mass varying with the square of the pion mass, as anticipated from the Goldstone nature of the pion. The results naturally follow those of the lattice for unphysically heavy pions~\cite{tandy-watson,benhaddou}, while allowing a natural bridge to the real pion mass, $m_\pi^2=0.02$ GeV$^2$. Considerable progress has been made with other quantum numbers too. These are in general more sensitive to details of the approximations beyond the rainbow-ladder for the quark-quark scattering kernel, e.g ~\cite{tandy-watson,williams-fischer,robertssquared}. While all the calculations discussed here are in Euclidean space, or equivalently spacelike momenta in Minkowski space, attempts to continue to timelike momenta have also been initiated~\cite{mink}. Similarly detailed dynamical calculations of the electromagnetic form-factors of the pion and nucleon have been made~\cite{cloet}. They are becoming more realistic and hold out the prospect that the momentum behaviour of the {\em up} and {\em down} mass functions shown in Fig.~4 may be amenable to experimental test with precision measurements of these form-factors~\cite{robertssquared}, planned for the 12 GeV upgrade at JLab.

\vspace{2mm}
Before rushing off and computing more complex hadronic effects from 
these basic quark elements, it is important to return to the 
behavour of the gluons and ghosts themselves that are essential 
for solving the quark SDE of Fig.~1. In fact, this was the 
subject of an exciting and excitable parallel session 
at this workshop. As already mentioned, in the pioneering  
Alkofer, von Smekal, Fischer {\em et al.} treatment in the 
Landau gauge~\cite{smekal,fischer-phd,fa}, the gluon is 
suppressed in the infrared, while the ghost is singular, Fig.~3.
Importantly, such low momentum behaviour is supported by 
studies~\cite{rgf} of Renormalization Group flow.
A singular ghost propagator means it dominates almost everywhere 
it can appear, as discussed by Schwenzer {\em et al.}~\cite{schwenzer}.
  According to~\cite{llanes-estrada}, the infrared dominance of the 
ghost is the key to confinement. 
At its simplest, confinement can be studied non-relativistically 
by considering infinitely heavy quarks. Then the inter-quark potential 
is dominated by one (dressed) gluon exchange with bare vertices. 
If  this were the case then an infrared vanishing gluon would not 
produce a rising potential at large distances. The vertices are expected to be bare because any corrections 
introduce additional quark lines and if these have infinite mass 
they are suppressed. 
The successes of Heavy Quark Effective Theory (HQET) rely on this property.  It is here that the singular ghost feeds in~\cite{llanes-estrada}. The vertex correction with a ghost loop brings  an infrared singular enhancement to the bare quark-gluon vertex, sufficient to generate a confining potential. This however comes at a price, the mass function for quarks like the $b$ have to behave quite differently from that for the {\em up} and {\em down} quarks. While  the $b$ quark is 5 GeV at short distances, they find (as shown in the lower part of Fig.~5) its mass function is less than 1~GeV  at 1~fm. This disagrees with the notions that underlie HQET. 

\vspace{3mm}
\begin{figure}[bh]
  \includegraphics[width=0.80\columnwidth]{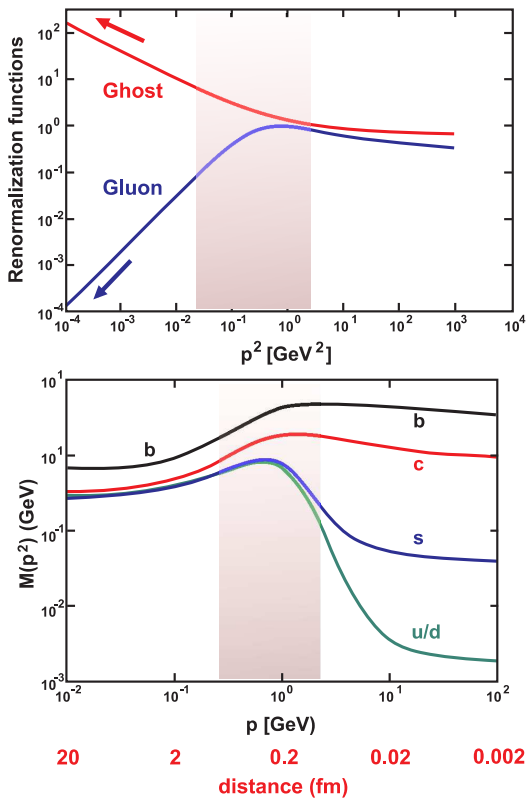}
  \caption{Log-log plots of (a) the gluon and ghost renormalization functions, and (b) the quark mass functions from the studies in Refs.~\cite{fischer-phd,llanes-estrada}. The shaded band marks the momentum region that largely controls hadron physics.}    
\end{figure}

One may question~\cite{mrp-eft09} whether this behaviour seen in the lower graph of Fig.~5 
is just an artefact of the approximations and truncations made. 
Let us follow the approach in~\cite{fischer-phd} and 
examine the coupled gluon and ghost propagator equations, 
where the interactions, the vertices, are assumed to be proportional 
to their bare structures.  As discussed above in the context of 
the photon in QED, Eq.~(1), the vector boson propagator 
depends on two functions $\,A\,$ and $\,B\,$ that gauge invariance 
requires to be equal.  Fischer {\em et al.}~\cite{fischer-phd,fa} 
solve the equations for $A(p^2)$ and this yields the scaling behaviour 
mentioned earlier with $\kappa\simeq 0.6\,$. However, explicit 
calculation shows (Fig.~6) this is not the same as $B(p^2)$. 
 Though the difference is less than a few percent at momenta 
below 1~GeV~\cite{wilson}, this makes a substantial difference 
to the existence of the solutions. Indeed, demanding these functions 
are equal, and the output gluon is transverse, there is no 
consistent value for $\kappa$ with the approximations used 
in~\cite{fischer-phd} as noted there.

 The key property of the gauge boson propagator is that it should be transverse in the Landau gauge both on the right and left hand sides of the gluon equation, as well as being multiplicatively renormalizable. This means it must not be more than logarithmically divergent in the ultraviolet. A non-perturbative truncation need not automatically fulfil this. The way to regulate the inverse propagator in the Landau gauge for a truncated system of Schwinger-Dyson equations is defined by
\be
\Pi^{\mu\nu}_{\mathrm{reg}}(p)\;=\;\Pi^{\mu\nu}(p)\;-\;\frac{g^{\mu\nu}}{p^2}\,p_\alpha\,\Pi^{\alpha\beta}(p)\,p_\beta\quad .
\ee
That this is an appropriate regularisation can be checked by considering the contribution of every graph at any order in perturbation theory. With dimensional regularisation, every graph will be divergent and non-transverse, but when all contributions at a given order are added they become purely logarithmically divergent in the ultraviolet and automatically transverse. The correct contribution of every graph to the total is ensured by the regularisation given in Eq.~(6).
This is equivalent to determining the gluon's behaviour from the function $\,B(p^2)\,$. As noted by Fischer~\cite{fischer-phd} with the simple interaction terms used here, a consistent scaling solution then requires $\,\kappa=1\,$. However,  such a solution, though valid in the infrared, has no connection in the ultraviolet to perturbation theory.  Motivated by the Renormalization Group arguments mentioned earlier~\cite{rgf}, Fischer {\em et al.}~\cite{cookery} have recently, within the SDE approach, constructed a more complicated dressing of the vertices, introducing factors that allow the infrared and ultraviolet behaviours to be less tightly linked through the shaded region of Fig.~5. Then an infrared scaling solution for both $A$ and $B$ functions of Eq.~(1) can be formed.  Nevertheless with a truncated SDE a consistently transverse gluon at all momenta requires a regulator like that of Eq.~(6). Moreover, as illustrated above even for  QED, this essential regulator does not ensure the boson propagator is physical --- recall Fig.~2. That requires appropriate interaction terms. Indeed, Bloch~\cite{bloch} has emphasised the key role played by the two loop graphs in the gluon SDE, which we ignore here, in this regard.

\begin{figure}[b]
  \includegraphics[width=0.95\columnwidth]{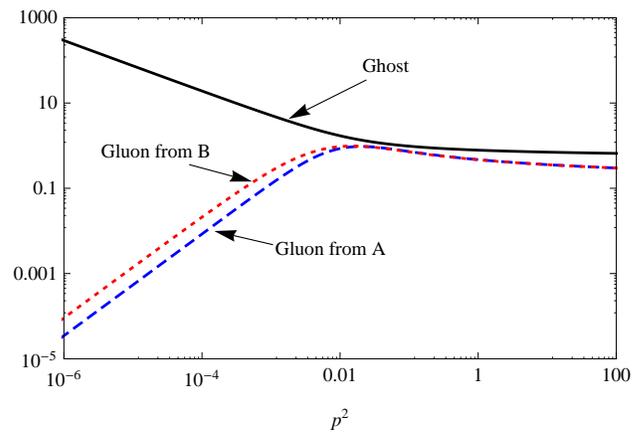}
  \caption{Log-log plot of the momentum dependence of the gluon and ghost renormalization functions from an analysis~\cite{wilson} like that of \cite{fischer-phd}. In the case of the gluon two versions are plotted:
short dash from $p^4/A(p^2)$ and the long dash from $p^4/B(p^2)$ of Eq.~(1).}
\end{figure} 

\begin{figure}[th]
  \includegraphics[width=0.89\columnwidth]{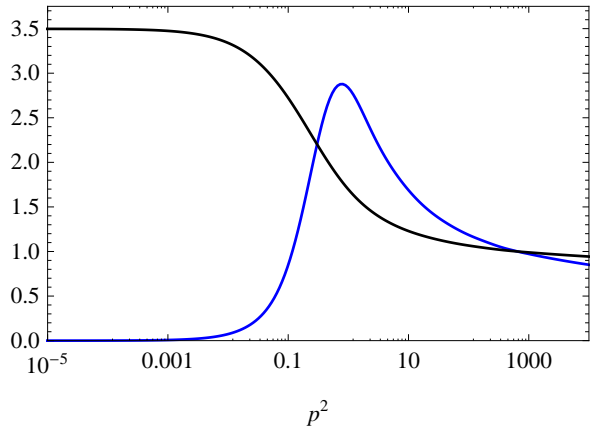}
  \caption{Linear-log plot of the momentum dependence of a self-consistent solution for the gluon and ghost renormalization functions with a finite value at $p^2=0$ for the ghost function, as described in the text~\cite{wilson}.}
\end{figure} 
 Meanwhile, the Orsay group of Pene {\em et al.}~\cite{pene} 
have long investigated the ghost propagator equation in some detail, 
and in particular, the structure of the ghost-gluon vertex for which 
Taylor proved a non-renormalization theorem~\cite{taylor,watson-thesis}. 
They have shown how with a vanishing gluon  (where $\kappa=0.5$ or a 
little bigger) the ghost renormalisation function tends to a constant 
in the infrared, as presented at this meeting by Rodriguez-Quintero~\cite{rod}.
 Calculations  with my collaborator David Wilson~\cite{wilson}, 
confirm the robustness of this conclusion. 
A \lq\lq scaling'' solution is not an automatic consequence of 
the Schwinger-Dyson approach, with its necessary truncations.

An infrared suppressed gluon more naturally leads to a ghost dressing function that is finite as $p\to 0$. But can such behaviour be found by consistently solving the gluon and ghost propagator equations simultaneously? The result is sensitive to the exact nature of the ghost-gluon vertex. If the ghost loop does dominate the equation for the inverse gluon propagator, as in the scaling solution, then Lerche and von Smekal~\cite{lerche} have suggested the ghost-gluon vertex itself might be transverse to ensure the gluon is 
consistently transverse.
 Of course, a transverse ghost-gluon vertex is too strong a requirement, difficult to reconcile with multi-loop orders in perturbation theory. The QED experience indicates that a simpler constraint is required. If one regularises the truncation of the gluon propagator equation using Eq.~(6), a self-consistent outcome for the dressing functions is shown in Fig.~7. The result is a \lq\lq massive" gluon solution, long advocated by Papavassiliou and collaborators~\cite{papa}, way back to Cornwall and Tiktopoulos~\cite{cornwall}. The solutions in Fig.~7 have been defined by momentum subtraction at $p=M_Z$, where the ghost and gluon renormalization functions are set equal to one. As discussed in~\cite{pene,wilson,rod}, such a subtraction precludes an infinite ghost at zero momentum. 
 The coupling used is equivalent to $\alpha_s (M_Z) = 0.118$ in the $\overline{MS}$ scheme. For such couplings, a massive gluon solution is thus most likely. However Fischer {\em et al.}~\cite{cookery} claim that a scaling solution, as favoured by Renormalization Group flow arguments~\cite{rgf} in the infrared, is still possible. So far implementing this within the SDE approach introduces ill-constrained vertex dressings~\cite{cookery}, indicating yet more work is required.

Of course,  one must remember that physics of the hadron world does not depend on the way ghosts and gluons propagate over distances of atomic scales, but only those of the size of a nucleus --- see Fig.~5 for the distance scale. Then the difference between a scaling and a massive gluon solution is far less marked. 
Nevertheless the dressed quark propagators, using the \lq\lq massive'' gluon of Fig.~7 with a non-singular interaction in the quark-gluon vertex of Fig.~1, do become like those expected of HQET. Moreover the ghost and gluon renormalization functions of Fig.~7 are in close accord with a multiplicity of recent lattice results~\cite{cucchieri,berlin}. We leave for others (in the parallel sessions) to report on  the finite volume and lattice spacing limitations of such studies.

What is more the Green's functions of Figs.~7,~4 can readily be input into bound state equations to predict the electromagnetic form-factors and electro-couplings of excited N$^*$'s that are of such physical interest~\cite{cloet,robertssquared}.
Both the SDE-BSE approach in the continuum and lattice studies connect quarks,
 gluons and ghosts to the static and dynamical properties of hadrons.  In the lattice approach this connection is a ``black box''. However, with the SDE-BSEs it is far less opaque, and we can see the inter-related mechanisms at work. It is this that makes this a worthwhile long term study.
Considerable progress has already been made. More is to come.

\vspace{-2mm}

\begin{acknowledgments}

\noindent It is a pleasure to thank Jos\'e Pelaez and Felipe Llanes-Estrada for inviting me to this instructive meeting.
This work was supported in part by the EU-RTN programme, Contract No. MRTN--CT-2006-035482, \lq\lq Flavianet''.
The report was authored  by Jefferson Science Associates, LLC under U.S. DOE Contract No. DE-AC05-06OR23177.
\end{acknowledgments}


\end{document}